\documentclass[sigconf,screen]{acmart}

\usepackage{soul}
\usepackage{xspace}
\usepackage{url}
\usepackage{graphicx}
\usepackage{listings,multicol}
\usepackage[caption=false]{subfig}  
\usepackage[export]{adjustbox}
\usepackage{amsmath}
\usepackage{textcomp}
\usepackage{breakurl} 
\usepackage{listings}
\usepackage{cleveref}
\usepackage{algorithm}
\usepackage{enumitem}
\usepackage[noend]{algpseudocode}

\newcommand{\todo}[1]{}

\newcommand{\ie}{\emph{i.e.,}\xspace}
\newcommand{\eg}{\emph{e.g.,}\xspace}

\algnewcommand\algorithmicforeach{\textbf{for each}}
\algdef{S}[FOR]{ForEach}[1]{\algorithmicforeach\ #1\ \algorithmicdo}
\newcommand{\vars}{\texttt}
\newcommand{\func}{\textrm}

\begin{document}

\title{Authorship Attribution of Source Code: A Language-Agnostic Approach and Applicability in Software Engineering}

\author{Egor Bogomolov}
\email{egor.bogomolov@jetbrains.com}
\affiliation{
    \institution{JetBrains Research}
    \institution{Higher School of Economics}
    \city{Saint Petersburg}
    \country{Russia}
}
\author{Vladimir Kovalenko}
\email{vladimir.kovalenko@jetbrains.com}
\affiliation{
    \institution{JetBrains Research}
    \institution{JetBrains N.V.}
    \city{Amsterdam}
    \country{The Netherlands}
}
\author{Yurii Rebryk}
\email{y.a.rebryk@gmail.com}
\affiliation{
    \institution{Higher School of Economics}
    \city{Saint Petersburg}
    \country{Russia}
}
\author{Alberto Bacchelli}
\email{bacchelli@ifi.uzh.ch}
\affiliation{
    \institution{University of Zurich}
    \city{Zurich}
    \country{Switzerland}
}
\author{Timofey Bryksin}
\email{timofey.bryksin@jetbrains.com}
\affiliation{
    \institution{JetBrains Research}
    \institution{Higher School of Economics}
    \city{Saint Petersburg}
    \country{Russia}
}


\begin{abstract}
Authorship attribution (\ie determining who is the author of a piece of source code) is an established research topic. State-of-the-art results for the authorship attribution problem look promising for the software engineering field, where they could be applied to detect plagiarized code and prevent legal issues.
With this article, we first introduce a new language-agnostic approach to authorship attribution of source code.
Then, we discuss limitations of existing synthetic datasets for authorship attribution, and propose a data collection approach that delivers datasets that better reflect aspects important for potential practical use in software engineering.
Finally, we demonstrate that high accuracy of authorship attribution models on existing datasets drastically drops when they are evaluated on more realistic data. We outline next steps for the design and evaluation of authorship attribution models that could bring the research efforts closer to practical use for software engineering.
\end{abstract}

\keywords{
Copyrights, Machine learning, Methods of data collection, Software process, Software maintenance, Security
}


\maketitle


%

\section{Introduction}\label{sec:introduction}

The task of source code authorship attribution can be formulated as follows: given a piece of code and a predefined set of authors, to attribute this piece to one of these authors, or judge that it was written by someone else. This problem has been of interest for researchers for at least three decades~\cite{Oman1989}. 

Prior research has shown that software engineering tasks, such as software maintenance~\cite{Anvik2006,Fritz2010,Girba2005} and software quality analysis~\cite{Bird2011,Thongtanunam2016,Rahman2011,Yin2011}, benefit from authorship information. Since authorship information in software repositories may be missing or inaccurate (\eg due to pair programming, co-authored commits, and changes made after code review suggestions), authorship attribution techniques could help in these tasks. Authorship attribution models output a probability for each of the known developers to be the author of a code fragment. When the probabilities are significantly higher for a group of developers, they are likely to be co-authors of the snippet, even if this information is missing in the version control system (VCS). On the other hand, when the commit was authored by several people but probability is high only for a single person, we can derive the main author of the commit.

Source code authorship attribution is also useful for plagiarism detection, either to directly determine the author of plagiarized code~\cite{Zhang2017,Burrows2007,Kothari2007} or to ensure that several fragments of code were written by a single author~\cite{Stein2011}. Importantly, while plagiarism detection techniques need a large database of code as a reference when checking for plagiarism, authorship attribution approaches can directly identify suspicious code snippets. 
Plagiarism detection, in turn, is important in software engineering: Software companies need to pay extra attention to copyright and licensing issues, as they can become liable to lawsuits~\cite{GoogleOracle}. For example, developers often use Stack Overflow~\cite{wu2019developers} to copy and paste code snippets to their projects. However, if developers do not use caution, code borrowed from Stack Overflow can induce licence conflicts on top of complicating maintenance~\cite{Baltes2019,Golubev2020}. 

Recently, several pieces of work improved the state of the art in authorship attribution on datasets for three popular programming languages: C++, Python, and Java. For C++, Caliskan \emph{et al.} reported the accuracy value of 92.8\% when distinguishing among 1,600 potential code authors~\cite{Caliskan2015}. For Python, Alsulami \emph{et al.} attributed code of 70 programmers with 88.9\% accuracy~\cite{Alsulami2017}. Yang \emph{et al.} developed a neural network model that achieved 91.1\% accuracy for a dataset of Java code by 40 authors~\cite{Yang2017}.

Some of the existing approaches use language-specific features. While language-specific features can improve the accuracy of authorship attribution for a particular language, designing a set of such features is a complex manual task. To transfer an approach that targets a specific language to another one (\eg from C++ to Python), one either needs to come up with a new set of features, or otherwise suffer from a major accuracy drop~\cite{Caliskan2015}.

In this study, we achieve higher results in authorship attribution accuracy by suggesting two \emph{language-agnostic} models. Both models work with path-based representations of code~\cite{Alon2018}, which can be built based on an abstract syntax tree (AST) for any programming language. The first model, called PbRF (Path-based Random Forest), is a random forest trained on relative term frequencies of tokens and paths in ASTs. To the best of our knowledge, PbRF is the first model that integrates path-based representations into a classical ML model. The second model, named PbNN (Path-based Neural Network), is an adapted version of the \emph{code2vec} neural network~\cite{Alon2018}. 

We evaluate both models on the datasets of code in three programming languages that were used in previous work. PbRF improves the state of the art for Java, C++, and Python datasets, even with few available samples per author. PbNN outperforms PbRF when the number of available samples per author is large. Both models improve state-of-the-art results for Java on a dataset of Yang \emph{et al.}~\cite{Yang2017}, with 98.5\% and 97.9\% accuracy respectively. Later, we evaluate how both models perform on our new collected datasets.
 
Another aspect that we target in this study is data used for evaluation. Existing work on authorship attribution operates with sources of data different from regular software projects: examples from books~\cite{Oman1989,Frantzeskou2006}, students' assignments~\cite{Elenbogen2008,Frantzeskou2007,Krsul1997}, solutions of programming competitions~\cite{Caliskan2015,Rosenblum2011-2,Alsulami2017,Simko2018}, and open-source projects with a single author~\cite{Kothari2007,Shevertalov2009,Yang2017,Zhang2017}. This data is different from code that can be found in real-world software projects. In this study, we deeply investigate this difference. Based on the results of this investigation, we propose a new data collection technique that can reduce these differences and generate more realistic datasets for the authorship attribution task. To formalize the differences, we suggest the concept of \textit{work context}: \ie aspects that can affect developer's coding practices and are specific to the concrete project, such as project's domain, team, internal coding conventions, and more. Also, we discuss another source of data differences: the evolution of programmers' individual coding practices over time and changing context of their contributions.

We propose a method to quantitatively measure the impact of work context and evolution of individual coding practices on the accuracy of authorship attribution models. Our evaluation shows that the accuracy of authorship attribution models plunges when models are tested with more realistic data than what is offered by existing datasets. In particular, the model that can distinguish between 40 authors with 98\% accuracy in one setup, reaches only 22\% for 21 developers in another. This result suggests that---before their practical adoption for software engineering---existing results in the field of source code authorship attribution should be revisited to evaluate robustness of the models and their applicability to more realistic datasets.

We made the artifacts related to this work publicly available on GitHub~\cite{JbrAuthorshipDetection} under MIT License. They contain implementation of the models, code for running the experiments, and a tool for collection of datasets for the authorship attribution task.

\smallskip

\noindent With this work we make the following contributions:

\begin{itemize}
    \item Two language-agnostic models that work with path-based representation of code. Both models can take as an input any code fragment (\eg file, class, method). These models outperform language-specific state-of-the-art models on existing datasets.
    \item An in-depth discussion on the limitations of existing datasets, supported by quantitative evaluation of effects of these limitations, particularly when applied to the software engineering domain.
    \item A novel, scalable approach to data collection for evaluation of source code authorship attribution models.
    \item A discussion of the concept of \textit{developer's work context} and a novel methodology to evaluate its influence on the accuracy of authorship attribution models.
    \item Empirical evidence on how the evolution of developers' coding practices impacts the accuracy of current authorship attribution models.
\end{itemize}






\section{Background}\label{sec:background}

To the best of our knowledge, the first work on source code authorship attribution dates back to Oman \emph{et al.}~\cite{Oman1989} in 1989. Although the results and approaches for the authorship attribution have changed and improved since then, the formulation of the problem did not change, and the underlying idea remains to use machine learning based on the features extracted from source code.

In the task of source code authorship attribution, a model is given a piece of code, and it should attribute the piece of code to one of the known developers, or state that it was written by someone else. The input code can differ by its form (the code can be unchanged, obfuscated, or even decompiled) and source of origin. In previous work, four sources of data have been used: 
\begin{description}[leftmargin=0.3cm]
    \item[\textbf{Code examples from books}] (\hspace{1sp}\cite{Oman1989,Frantzeskou2006}). Used before the era of easily available open-source projects, for the lack of other sources.
    \item[\textbf{Students' assignments}] (\hspace{1sp}\cite{Elenbogen2008,Frantzeskou2007,Krsul1997,Burrows2014}). Often, researchers are not allowed to publish these datasets (\eg from university courses) due to privacy or intellectual property issues. Lack of published data causes problems comparing different methods.
    \item \textbf{Solutions to programming competitions} (\hspace{1sp}\cite{Caliskan2015,Rosenblum2011-2,Alsulami2017,Simko2018}). This mostly refers to data from Google Code Jam\footnote{\url{https://codingcompetitions.withgoogle.com/codejam}} (GCJ), an annual competition held by Google since 2008.
    \item[\textbf{Single-author open-source projects}] (\hspace{1sp}\cite{Kothari2007,Shevertalov2009,Yang2017,Zhang2017}). With the increasing popularity of hosting platforms for open source projects (\eg GitHub), they have become a major source of data. In this case, the dataset consists of multiple repositories, each developed by a single programmer. For evaluation, the researchers use unseen code snippets from the same repositories. Researchers avoid repositories with multiple authors because in this case authorship of even small fragments of code might be shared.
\end{description}

According to the recent survey by Kalgutkar \emph{et al.}~\cite{Kalgutkar2019}, the following are the best results per programming language:
\begin{description}[leftmargin=0.3cm]
    \item[\textbf{C++}]: Caliskan \emph{et al.}~\cite{Caliskan2015} reported the best results using a random forest trained on syntactical features. They achieved 93\% and 98\% accuracy for datasets with 1,600 and 250 developers, respectively.
    \item[\textbf{Python}]: Alsulami \emph{et al.}~\cite{Alsulami2017} suggested to use tree-based LSTM to derive implicit features from ASTs, achieving 88.9\% accuracy in distinguishing among 70 authors.
    \item[\textbf{Java}]: Yang \emph{et al.}~\cite{Yang2017} reported 91\% accuracy for a dataset of 40 authors using neural networks. Instead of a commonly used stochastic gradient descent optimizer, the authors trained the network with particle swarm optimization~\cite{Kennedy1995} improving the accuracy by 15 percent points.
\end{description}

Syntactic features, derived from AST of code, are known to improve the results for authorship attribution~\cite{Caliskan2015,Alsulami2017} as well as for other software engineering tasks, such as code summarization~\cite{AlonCode2Seq}, method name prediction~\cite{Alon2018}, and clone detection~\cite{Perez2019}.

\smallskip
Compared to real-world data, where a programmer often works on multiple projects and using multiple languages, existing datasets are limited to a single language and one project per author.
To overcome this limitation, models applicable in real-world environment should work with different programming languages in a consistent manner. Following this idea, we decided to build a language-independent model, that is based on syntactic features and works on par with prior studies.

\section{Language-Agnostic Models}\label{sec:models}

Our first goal is to develop an authorship attribution solution that is language-agnostic and achieves an accuracy comparable to state-of-the-art approaches.

To apply machine learning methods to code, one should transform code into a numerical form called \emph{representation}. While some works use explicitly designed language-specific features~\cite{Caliskan2015,Yang2017}, we represent code using \emph{path-based representation}~\cite{Alon2019} to work with code in various programming languages in a uniform way.

A common way to use path-based representation is the \emph{code2vec} neural network~\cite{Alon2018}, suggested by Alon \emph{et al.} for the task of method name prediction. However, code2vec requires a significant number of samples for each author to infer meaningful information, due to the large number of trainable parameters.
Thus, alongside with the neural network, we also employ a random forest model,  trained on similar features.
The random forest model shows a better accuracy for small datasets, but generalizes worse for the larger ones. In the rest of this section, we describe our models and define related concepts.

\subsection{Definitions}

\begin{figure}[htp]
    \centering
    \subfloat[An example code fragment]{
        \includegraphics[clip,width=0.4\columnwidth]{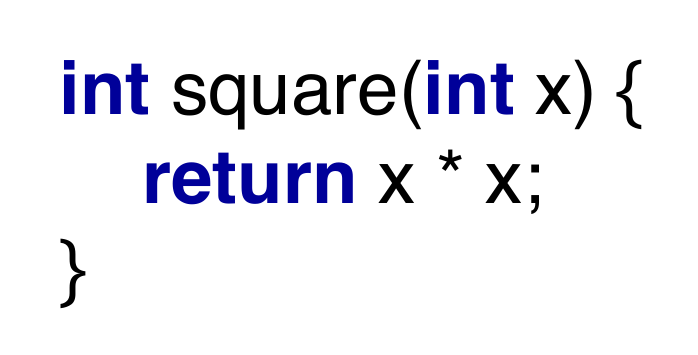}
        \label{fig:ast-code} 
    }
    
    \subfloat[An AST of this code fragment]{
      \includegraphics[clip,width=\columnwidth]{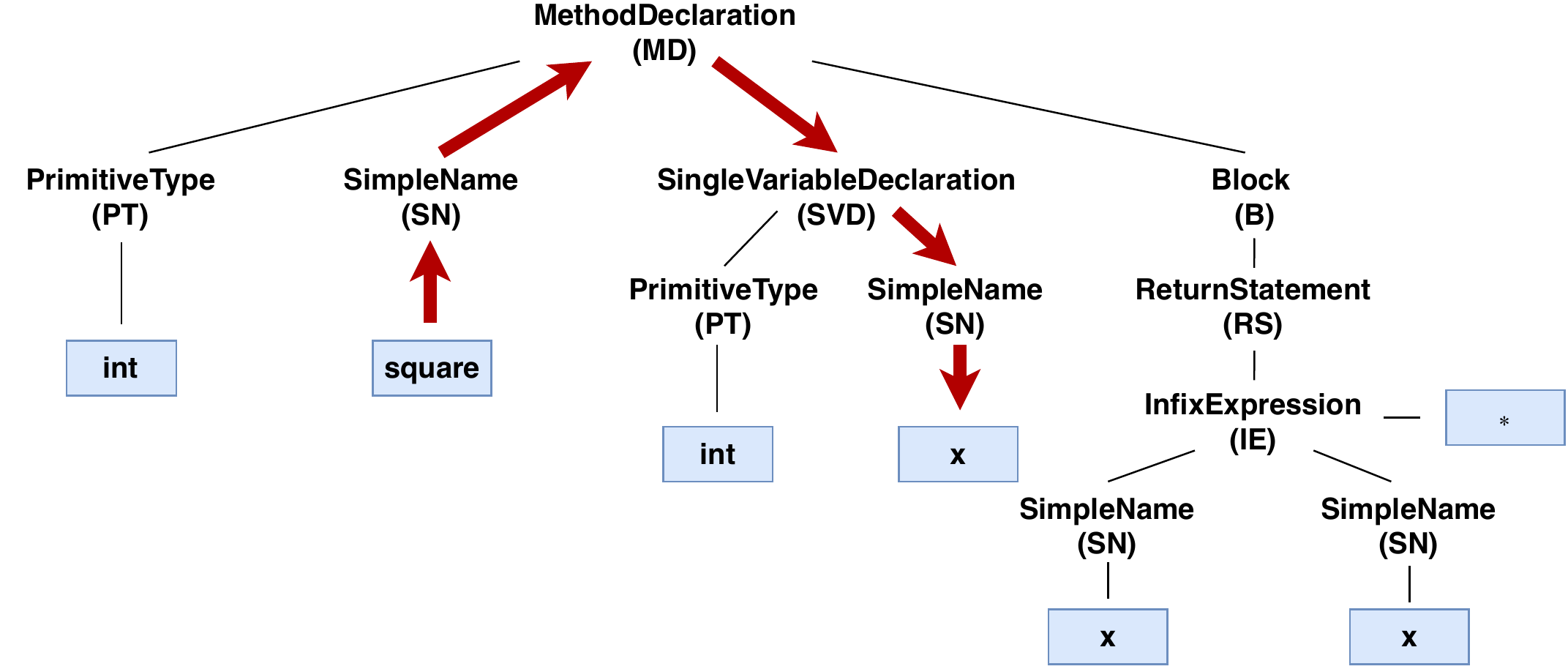}
      \label{fig:ast-example}
    }
    \centering
    \caption{A code example and a corresponding AST}
    \label{fig:ast}
\end{figure}

\noindent\textbf{Abstract Syntax Tree.}
An abstract syntax tree (AST) is a representation of program's code in the form of a rooted tree. Nodes of an AST correspond to different code constructs (\eg math operations and variable declarations). Children of a node correspond to smaller constructs that comprise its corresponding code. Different constructs are represented with different \textit{node types}.
An AST might omit parentheses, tabs, and other formatting details. 
\Cref{fig:ast} shows an example of a code fragment and a corresponding AST.

\noindent\textbf{Path in AST.}
A \textit{path} is a sequence of connected nodes in an AST. Start and end nodes of a path may be arbitrary, but we only use paths between two leaves in the AST to conform with code2vec and have the benefit of working with smaller pieces of code that such paths represent. 
Following Alon \emph{et al.}~\cite{Alon2018}, we denote a path by a sequence of node types and directions (up or down) between consequent nodes. The node in which the path changes direction is called \emph{top node}. In \Cref{fig:ast-example}, an example of a path between the leaves of an AST is shown with red arrows. In the notation of node types and directions, this path can be denoted as follows:
$$
SN \uparrow MD \downarrow SVD \downarrow SN
$$

\noindent\textbf{Path-context.}
Path-contexts are triplets, consisting of a path between two leaves and tokens corresponding to start and end leaves.
A path-context represents two code tokens and a structural connection between them. This allows a path-context to capture information about the structure of code. Prior works show that code structure also carries semantic information~\cite{Alon2018,AlonCode2Seq}.
\Cref{fig:ast-example} highlights the following path-context:
$$
(square, SN \uparrow MD \downarrow SVD \downarrow SN, x)
$$

This path-context represents a declaration of a function named \emph{square} with a single argument named \emph{x}. 
The path in this path-context encodes the following information: It contains nodes \textit{Function Declaration} as well as \textit{Single Variable Declaration} and tokens are linked to \textit{Simple Name} AST nodes.

\noindent\textbf{Path-based representation.}
A path-based representation treats a piece of code as a bag of path-contexts. For larger pieces of code, the number of path-contexts in the bag might be large and can be reduced by setting a limit on the length (\ie the number of vertices in the list) and width (\ie the difference in indices between the children of the top node) of the paths. These limits on the length and width are hyperparameters and are determined empirically. For each piece of code, we generate all path-contexts that satisfy the limits on the length and width. However, we use a random subset of mined path-contexts to train the models, as explained later in this section.

To adjust a bag of path-contexts comprising the path-based representation to the task of training a model, we need to transform it into a numerical form. For the code2vec model, an embedding layer translates paths and tokens into numerical vectors. For the random forest model, we cannot use embedding because random forest trains without gradient computation. In this case, the transformation is done by computing relative term frequencies of paths and tokens (\ie the number of times the token/path occurs in a code snippet divided by the total number of tokens/paths) and storing them in a sparse vector. Since the path-based representation does not require any specific properties from the programming language, both PbRF and PbNN are language-agnostic. The following subsections cover both cases in more detail.

\subsection{PbRF (Random Forest Model)}\label{subsec:pbrf}

The random forest model is designed to work even when the number of samples for each author is rather small (starting from a few samples per author), where the neural network cannot capture enough information to generalize. Random forest has already proved to be effective in this setup in previous work~\cite{Caliskan2015,Simko2018}.

Tokens and AST-based features have already proved to be effective for authorship attribution in the work by Caliskan \emph{et al.}~\cite{Caliskan2015}. Compared to their work, we use more complex AST features: AST paths. Random forest does not allow training an embedding of path-contexts, thus, instead of combining paths and tokens into path-contexts, we use relative term frequencies of tokens and paths as features. To the best of our knowledge, this work is the first to combine path-based representations with classical machine learning models (such as Random Forest).

If a set of documents contains $T$ tokens and $P$ paths, the random forest model takes a sparse vector of size $F=P+T$ as an input. The size of such a vector might be significant (up to millions), with some features being unimportant for identifying the author. We employ feature filtering to reduce the effect of this dimensionality problem. As in previous work on authorship attribution~\cite{Rosenblum2011, Meng2016, Caliskan2015}, we use filtering based on \emph{mutual information} (MI)~\cite{elementsOfInfoTheory}. The mutual information of a source code feature $f$ and an author $A$ can be expressed as:
$$MI(A, f) = H(A) - H(A|f),$$
where $H(X)$ is Shannon entropy~\cite{Shannon1948}. For the task of authorship identification, we interpret it as follows: the higher the MI is (\ie the lower $H(A|f)$ is), the better one can recognize the author based on the value of the given feature. 

Feature selection based on the mutual information criteria ranks all the features by their MI with the author label and takes $N$ with the highest MI value. $N$ is a hyperparameter determined empirically during the evaluation process by trying various values. This approach does not account for dependencies among features; for example, if there are two identical high-ranked features, we take both and miss some other feature. To avoid this problem, one could add features one by one and compute mutual information after each step, but on the scale of millions of features this procedure is too costly.



\subsection{PbNN (Neural Network Model)}\label{subsec:pbnn}


To achieve a better accuracy for larger datasets, we adopted the neural network called code2vec~\cite{Alon2019}. It takes a bag of path-contexts from a code snippet as an input, transforms them into a single numerical vector, and predicts a probability for each known developer to be the author of the snippet. Compared to classical machine learning methods, neural networks can derive more complex concepts and relationships from structured data, when given enough training samples. A detailed description of the model is available in the original \emph{code2vec} paper by Alon \emph{et al.}~\cite{Alon2019}.



The number of the PbNN's parameters is $O((T + P)d)$. Since the value of $(T+P)$ is usually large, from tens of thousands to millions, the number of required samples for the model to train is also significant.

\section{Evaluation on Existing Datasets}\label{sec:evaluation-comparison}

We evaluated the two models presented in the previous section on the publicly available datasets for Java, C++, and Python used in recent work~\cite{Caliskan2015,Alsulami2017,Yang2017} and compared the accuracy of PbRF and PbNN to the results reported in these papers. \Cref{tab:dataset-stats} shows statistical information about the datasets.
\Cref{tab:results-comparison} presents the accuracy results of evaluation.

To compare the results of different models when these results are obtained through multiple runs (\ie folds in cross-validation), we apply the Wilcoxon signed-rank test~\cite{Wilcoxon} to various accuracy values per run. When only the mean accuracy is available (which is the case for previous work), or the number of runs is too small, we directly compare the mean values.

\begin{table}[ht]
\caption{Datasets used in previous works. The number of paths is provided for $width <= 2$ and $length <= 8$.}
\begin{center}
\begin{tabular}{lrrr}
    & C++~\cite{Caliskan2015} 
    & Python~\cite{Alsulami2017}
    & Java~\cite{Yang2017}\\ \hline
    Number of authors & 1,600 & 70 & 40 \\ 
    Number of files & 14,400 & 700 & 3,021 \\ 
    Samples per author & 9 & 10 & 11 to 712 \\
    Source & GCJ & GCJ & GitHub  \\ 
    Unique tokens & 50,000 & 10,000 & 65,200 \\ 
    Unique paths & 102,000 & 22,100 & 220,000
\end{tabular}
\end{center}
\label{tab:dataset-stats}
\end{table}

\begin{table}[ht]
\caption{Mean accuracy by approach and dataset.}

\begin{center}
\begin{tabular}{lrrr}
    
    & C++~\cite{Caliskan2015} & Python~\cite{Alsulami2017} & Java~\cite{Yang2017} \\ \hline
    
    Caliskan \emph{et al.}~\cite{Caliskan2015} & 0.928 & 0.729 & N/A \\ 
    Alsulami \emph{et al.}~\cite{Alsulami2017} & N/A & 0.889 & N/A \\ 
    SGD~\cite{Yang2017} & N/A & N/A & 0.760 \\ 
    PSO~\cite{Yang2017} & N/A & N/A & 0.911 \\ 
    \textbf{This work, PbNN} & 0.793 & 0.723 & \textbf{0.979} \\ 
    \textbf{This work, PbRF} & \textbf{0.948} & \textbf{0.959} & \textbf{0.985} 
\end{tabular}
\end{center}
\label{tab:results-comparison}
\end{table}

\subsection{Hyperparameters}
Both our models have parameters that should be fixed before the training phase, \ie hyperparameters. These hyperparameters are the number of trees, number of features left after feature selection, maximum depth of the trees for the random forest (\Cref{subsec:pbrf}), size of the embedding vector for the neural network (\Cref{subsec:pbnn}), and the limits on length and width of AST paths in path-based representation. We tuned these hyperparameters using grid search~\cite{ScikitLearn2011}. The plots of the models' accuracy dependence on the specific parameters can be found in supplementary materials.

We found that the optimal values are similar across the datasets. For the random forest model, increasing the number of trees improves the accuracy until the number reaches 300, after that the accuracy reaches a plateau. Increasing the maximum tree depth leads to accuracy growth and does not cause overfitting, so we removed the limits on the tree depth. The optimal number of features left after the selection is about 7\% of the initial number of features for all datasets. While a slightly different number of features may result in a higher accuracy, the differences are within the standard deviation range.

For the neural network, increasing the embedding size from 32 to 256 results in a significant growth in accuracy for the Python and C++ datasets. For the Java dataset, we did not notice any significant change, even though a larger size does not cause overfitting.

The models show a better accuracy when the limits on the length of paths in an AST are smaller. We tried the lengths from 6 to 10, and the accuracy is the best for 6 or 7. Short paths correspond to small repetitive constructs in code, which can be typical of a particular developer. Paths of a higher length are less frequent and cause the model to overfit. The same applies to the limits on the path width: increasing the width beyond 2 did not improve the attribution accuracy.

\subsection{Evaluation on C++}

The C++ dataset was introduced by Caliskan \emph{et al.}~\cite{Caliskan2015}. It contains solutions by each of 1,600 participants for 9 problems from Google Code Jam competitions from 2008 to 2014, making it one of the largest experiments with respect to the number of authors. Each author has 9 samples in the dataset, corresponding to the solutions of 9 problems. Following the original paper, we run a 9-fold cross-validation. At each fold, the held-out set contains a single solution from each author, while the training set contains 8 problems from each author. For comparison, we use the average models' accuracy over all folds.

Caliskan \emph{et al.}~\cite{Caliskan2015} reported 92.8\% mean accuracy after 9-fold cross-validation. PbNN and PbRF achieve 79.3\% and 94.8\% average accuracy, respectively. The neural model shows a lower accuracy because of overfitting: the number of available data points is too small to train a much larger set of the network's internal parameters. The PbRF's accuracy is higher compared to the Caliskan's work (94.8\% vs 92.8\%), and the difference lies out of the standard deviation range computed based on the cross-validation (which is 0.4\%). We can conclude that \emph{PbRF achieves a better accuracy than the previous best result}.

\subsection{Evaluation on Python}

The Python dataset also contains Google Code Jam solutions. It was collected and introduced by Alsulami \emph{et al.}~\cite{Alsulami2017} and consists of solutions to 10 problems implemented by 70 authors. As in the C++ dataset, this one contains the same number of samples for each author. During cross-validation, the model is trained on 8 problems and validated on 2 other problems that are initially held out. The best reported average accuracy is 88.9\%.

For comparison, we use the average accuracy reported by Alsulami \emph{et al.}~\cite{Alsulami2017} for two models. The first one is a novel model introduced by the authors. The second one is an adopted version of the model by Caliskan \emph{et al.}~\cite{Caliskan2015} with C++-specific features removed.
On this dataset, our models achieve 72.3\% (PbNN) and 95.9\% (PbRF) accuracy. Similarly to the C++ dataset, PbRF shows a better accuracy compared to PbNN because the number of available samples is too small for efficient training of the neural network. Also in this case, \emph{PbRF achieves an accuracy that is as good as the previous best result}.

\subsection{Evaluation on Java}

The Java dataset introduced by Yang \emph{et al.}~\cite{Yang2017} consists of 40 open source projects, each authored exclusively by a single developer. Each project contains from 11 to 712 files with a median value of 54, totaling 3,021 files overall. The number of samples per author varies.

For evaluation, we split the dataset into 10 folds and perform a stratified cross-validation similarly to Yang \emph{et al.}~\cite{Yang2017}. Ideally, to compare the accuracy of our model to theirs in the most precise manner, we would use an identical split of the dataset into folds for our evaluation. However, the original split into folds is not available and we created our own with a fixed random seed.

The model by Yang \emph{et al.} is a neural network trained either by stochastic gradient descent (SGD) or particle swarm optimization (PSO)~\cite{Kennedy1995}. When trained by PSO, the model achieves an average accuracy of 91.1\% using 10-fold cross-validation. Both our models reach an accuracy of more than 97\%. The previous result lies out of the standard deviation range in both cases, thus indicating that the difference is statistically significant. Although the median number of samples per author is only 54, the neural network shows a high accuracy.
Even though the average accuracy of the neural network model is slightly higher than that of the random forest, the statistical test yields a p-value of 0.07. Also in this case, both \emph{PbRF and PbNN improve the previous best result}.

\section{Limitations of Current Evaluations}\label{sec:discussion-threats}

During the analysis presented in \Cref{sec:evaluation-comparison}, we realized a number of limitations posed by this evaluation technique.
Particularly, using a single accuracy metric to compare complex approaches to problems motivated by practical needs (such as authorship attribution in software engineering) is a one-sided solution. In this section, we discuss the limitations of such evaluation.

Academic work on authorship attribution is motivated by practical needs such as detection of plagiarism~\cite{Ottenstein1976,Liu2006,Shevertalov2009}, detection of ghostwriting~\cite{Elenbogen2008,Kothari2007}, and attribution of malware~\cite{Layton2010,Caliskan2015,Frantzeskou2007,Krsul1997}. 
In a perfect world, introduced authorship attribution approaches should be evaluated on real-world data. However, such real-world data is privacy-sensitive and seldom publicly available. 
For this reason, to show how models behave and compare against each other, researchers create datasets from available data sources. Even though these datasets try to mimic the data found in practical applications, there exist major differences. To illustrate them, we discuss the concept of developer's \textit{work context}, \ie the environment that surrounds programmers when they write code. The work context includes but is not limited to the following:

\begin{itemize}
    \item \textbf{Parts of a codebase}: A codebase of a software project usually contains logically interconnected code that is organized into packages and modules. This logical connection often implies a lower-level connection observed in code: calls of methods, creation of objects, similar names of entities.
    \item \textbf{Project domains}: The domain of the task (\eg an Android application) influences names, used libraries, implemented features, and architectural patterns that are used in such projects more commonly.
    \item \textbf{Projects}: The project itself might have internal naming conventions and utility components that are called from different parts of the codebase. Moreover, some companies have their own style guides for programming languages that affect naming conventions, formatting, and preferred use of specific language constructions. An example of this is Google Style Guides~\cite{GoogleStyleGuides}.
    \item \textbf{Set of tools}: Integrated development environments (IDE) or text editors, version control systems, as well as build and deployment tools may influence the way how developers write code. For example, a recent survey of programmers concluded that developers affect their development practices by simply using GitHub in their projects~\cite{Kalliamvakou2015}.
\end{itemize}

This list is not complete and could easily be extended, but the effect of even a single of the aforementioned individual examples of work context might be significant for the task of authorship attribution. When collecting data for evaluation, we should take work context into consideration: practical applications of authorship attribution often imply that the model should be trained on code written in one context and tested in another, or distinguish between developers working in the same context. However, datasets used in prior works do the opposite: there is a difference between authors' work contexts (\eg different single-authored projects) and no difference between work contexts of the same developer in training and evaluation sets.

Another concern is that existing datasets do not consider the impact of collaboration on the code. All of them consist of projects developed by a single programmer. However, projects studied in the software engineering domain are usually developed by teams. Collaborative work is not reflected in prior work, but it may introduce additional complexity for the authorship attribution task: Developers integrate their code into codebases written in collaboration with their colleagues, which might make their code harder to distinguish.


Developers' individual coding practices may change over time~\cite{Burrows2009}. It is reasonable to think that changes in coding practices (\eg programmer's style, used libraries, naming conventions, development process) can influence the accuracy of the models significantly. For existing datasets, all code written by a single author belongs to roughly the same period of time (\eg one project, one competition, or assignments belonging to one course) and there is no temporal division between evaluation and training sets. Though, for practical problems, one might need to train a model on the historical data and apply it to new samples. This temporal aspect may introduce potential significant difference in individual coding practices between code used for training and testing.

Two prior studies consider evolution of programmers' style as a challenge for authorship attribution~\cite{Burrows2009,Caliskan2015}. Burrows \emph{et al.} evaluated the difference between six student assignments, showing that their coding style changes over time~\cite{Burrows2009}.
Caliskan \emph{et al.} trained a model on solutions of Google Code Jam (GCJ) 2012 and evaluated it on a single problem from GCJ 2014. Their experiment did not reveal any major differences in accuracy compared to evaluating on a problem from the same GCJ 2012~\cite{Caliskan2015}.
These results are contradictory, but both studies operated with small datasets and in domains different from real-world projects. Thus, further research on this topic is required.

\smallskip
We conclude that there is a gap (at least theoretical) between the existing datasets and what can be collected and used in the context of real-world applications. In particular, there is a difference in terms of work context, effects of developer collaboration, and changes over time. In the following sections, we suggest a novel approach to data collection that allows to quantitatively evaluate the impact of both temporal and contextual issues on the accuracy of authorship attribution models.

\section{Collecting Realistic Data}\label{sec:data}

To quantitatively evaluate the impact of dissimilarities between existing datasets and real software projects on the accuracy of authorship attribution techniques, we developed a new approach to data collection. It uses Git\footnote{https://git-scm.com/} repositories as the data source and, unlike existing datasets from open-source data~\cite{Lange2007,Yang2017,Alsulami2017,Zhang2017}, overcomes the limitation of a single author per project.

\subsection{Method of Data Collection}
We suggest a new approach to collecting evaluation data for authorship attribution models. The approach works with any Git project without restrictions on the number of developers. In particular, using Git as the main data source allows taking data from GitHub, the world's largest repository hosting platform with more than 100 million repositories and 30 million users~\cite{GithubStats}. Open-source Git repositories, and GitHub in particular, are a uniquely rich source of data for modern software engineering research efforts~\cite{Bird2009,Allamanis2014,Kalliamvakou2014,Menezes2018}.
The atomic unit of contribution in Git projects is a commit. Usually, a commit has a single author. This authorship information associated with every change recorded in a Git repository makes Git a particularly rich source of data for authorship attribution studies. 


The first step of our data extraction method is to traverse the history of a repository to gather individual commits. 
Then we need to identify commits authored by the same developer. It is not a trivial task, because a developer can work within one repository under multiple aliases using different emails. Even though there are prior studies on the problem of alias merging~\cite{Bird2006,Robles2005,Kouters2012}, these methods either make assumptions or are probabilistic to some extent. The main benefit of the existing alias merging methods is their complete autonomy, which allows working with arbitrary amounts of data. In this study, we processed the IntelliJ IDEA Community~\cite{IntelliJCommunity} repository by merging non-stub names and emails that appeared together into groups. Then we accessed GitHub to merge the groups that belong to the same person and contacted the developers to resolve all arguable cases. We took this approach since we processed a single repository and wanted to avoid any mistakes. For future researchers who may want to process larger amounts of data autonomously, our data preparation pipeline supports using existing entity-merging methods.




In the second step, we split each commit into changes of fixed granularity (\eg a change to a single class, method, or field). 
In this study, we use a \emph{method change} as the granularity unit: it is hard to precisely track changes of smaller fragments of code (\eg lines or statements), while using class-level or higher granularity would leave us with less data points. Also, authorship attribution at the method level can be a valuable task in the domain of clone detection, where a method is a commonly studied unit~\cite{Golubev2020}.

Within the scope of a single commit, methods can be renamed or moved to other files. We use GumTree~\cite{GumTree} to precisely track such changes in Java code as well as simple changes to a method's body.
As a result, we get a set of all method changes made during the project development. Afterwards, the extracted data can be grouped into datasets with different properties.

\subsection{Collected Datasets}

We implemented an open-source tool supporting the described approach to data collection which is available in the supplementary package. 
We used the tool to extract data from the IntelliJ IDEA Community Edition~\cite{IntelliJCommunity} project, the second largest public Java project on GitHub. At the time of processing (March 10th, 2020), the project contained about 270,000 commits dated from 2004 to 2020 and authored by 500 developers. These commits comprise about two million individual method changes. Each change is of one of three types: creation of a method, deletion, or modification of its body or signature. The latter, unlike method creations, cannot be processed directly by authorship identification models: newly added code fragments might be incomplete, and the concrete modifications might be scattered across the method body. Moreover, the author of the original code might be different from the one who modifies the method, which makes it impossible to define a sole author of the method.
In the datasets designed in this work, we only use method creations, because they contain new code fragments implemented by a single person and can be labeled accordingly. However, attributing authorship of method modifications is an interesting task for future research which will require changing the models' input (\eg take two versions of code as an input or process the difference between the versions).

In the IntelliJ IDEA Community repository, out of all the changes, 680,000 are method creations which is about a third of all the changes. Thus, authorship attribution models that operate only with method creations can cover a large part of all the cases. The developers highly differ by the amount of contributions: 20 most active developers created 58\% of all methods, 50 most active--- 88\%. To reduce the imbalance of the datasets, we split the authors in two groups: 21 authors with at least 10,000 created samples (\ie created methods) and 44 authors who have between 2,000 and 10,000 samples. In total, the groups have about 400,000 and 200,000 samples, respectively.

To quantitatively evaluate the impact of work context and evolution of coding practices on the quality of authorship attribution, we further create several datasets from the collected data. To make evaluation conditions as close to practical tasks as possible, we should have processed several projects and split \textit{projects} between training and testing sets. However, at this point it is unclear how to define the similarity between work contexts of different projects, and we would not be able to run several experiments with an increasing degree of context difference to perform a quantitative evaluation. Thus, this left us with one project and multiple datasets.

\subsubsection{\textbf{Datasets with gradual separation of work context}}\label{sec:dataset-context-separation}


The purpose of these datasets is to measure the influence of variation in developers' work context on the quality of authorship attribution. To achieve this, we need multiple pairs of training and evaluation samples that differ only in their work context. More specifically, the pairs should contain the same code fragments but be split differently between the training and testing part.

\begin{figure}[htp]
\centering
    \includegraphics[width=\columnwidth]{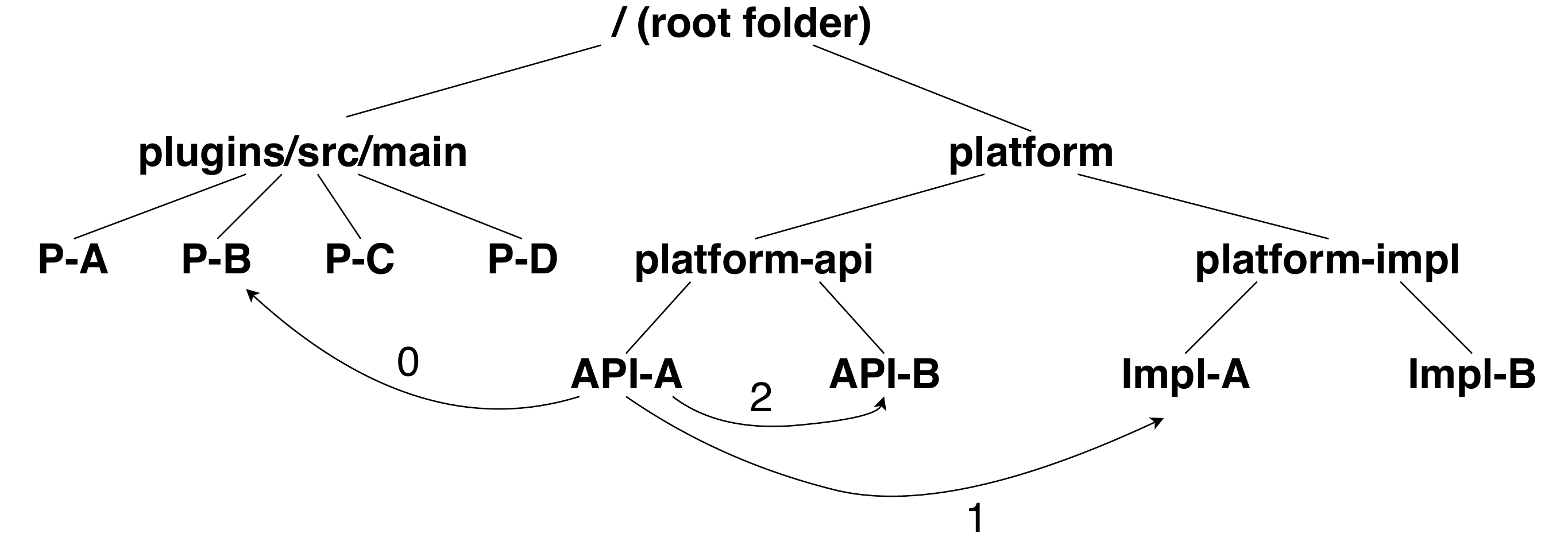}
    \centering
    \caption{An example of a project's file tree with similarities between files.}
    \label{fig:filetree}
\end{figure}

To control the degree of difference in work contexts, we use the project's file tree. \Cref{fig:filetree} shows an example of such a tree. It consists of folders with edges between a folder and its content. Leaves in the tree correspond to files. For Java code, we are interested only in Java source files identified by the `.java' extension. To reduce the depth of the tree, we compress paths of folders with a single sub-folder into nodes: In \Cref{fig:filetree}, folders ``plugins'', ``src'', and ``main'' are compressed into a single node ``plugins/src/main''. This operation preserves the structure of the tree.

\begin{algorithm}
    \caption{Finding split of a dataset with selected separation of work context.}\label{alg:context-separation}
    \begin{algorithmic}[1]
        \Function{RandomSplit}{$\vars{author},\vars{depth},\vars{testRatio}$}
            \State $\vars{folders} \gets \func{allFoldersAtDepth}(\vars{depth})$
            \State $\func{randomShuffle}(\vars{folders})$
            \State $\vars{train},\vars{test} \gets \vars{emptySet},\vars{emptySet}$
            \ForEach{$\vars{folder} \in \vars{folders}$}
                \State $\vars{samples} \gets \func{authorSamples}(\vars{author}, \vars{folder})$
                \If{$\func{getRatio}(\vars{train},\vars{test})<\vars{testRatio}$}
                    \State $\vars{test}.\func{insert}(\vars{samples})$
                \Else
                    \State $\vars{train}.\func{insert}(\vars{samples})$
                \EndIf
            \EndFor
            \State \Return{$\vars{train},\vars{test}$}
        \EndFunction
    
        \Function{SplitAuthor}{$\vars{author},\vars{depth},\vars{minR}, \vars{maxR},\vars{N}$}
            \State $\vars{testR} \gets \frac{\vars{minR}+\vars{maxR}}{2}$
            \State $\vars{bestSplit} = \vars{None}$
            \ForEach{$\vars{i} \in \func{range}(\vars{N})$}
                \State $\vars{split} \gets \func{RandomSplit}(\vars{author},\vars{depth},\vars{testR})$
                \State $\vars{ratio} \gets \func{getRatio}(\vars{split})$
                \If {$\vars{ratio} \in (\vars{minR}, \vars{maxR})$}
                    \State $\func{updateBest}(\vars{split},\vars{bestSplit},\vars{testR})$
                \EndIf
            \EndFor
            \State \Return{$\vars{bestSplit}$}
        \EndFunction
    \end{algorithmic}
\end{algorithm}

A file tree for a Java codebase resembles the structure of its packages. Usually, classes in one package are logically connected and refer to each other; thus, they have similar work contexts. At a higher level of abstraction, this also applies to classes in different sub-packages of the same package. In \Cref{fig:filetree}, the tree class \texttt{API-A} has a work context that is similar to \texttt{API-B}, because they are in the same package, but a less similar context to \texttt{Impl-A} from \texttt{platform-impl}. Nevertheless, they are still much closer to each other than to any file from the \texttt{plugins} package since both are used to implement some platform features and may even depend on one another.

From this observation, we derive a way to measure the similarity of work contexts of two files: it is \emph{the depth of the lowest common ancestor in the file tree}. In \Cref{fig:filetree}, similarities between \texttt{API-A} and other classes are shown with arrows (the depth of the root is considered to be 0). By doing so, for a training-validation split, we can define the similarity of work contexts of these parts as the highest value of pairwise similarity between files in them. Using the maximum value might seem over-cautious, but in real-world applications it is natural to assume that the model will not see any samples from another context during training before receiving them as input.

The subsequent step is to create a sequence of data splits with increasing similarity values, or the depth of split. \Cref{alg:context-separation} shows the pseudocode for creation of such splits. To preserve the distribution of authors at each level, we pick splits for different authors independently and merge them afterwards. We fix a ratio of evaluation samples, the depth of split, and an author. Then, we collect all folders at the chosen depth and files at this depth or higher. Afterward, we greedily divide them into training and evaluation parts, trying to get as close to the chosen ratio of testing samples as possible. When a folder is put into the training or evaluation part, all the methods created by the author in the subtree of the folder go into this part. We repeat the procedure several times for each author and depth, and pick the split with the ratio closest to the fixed one. If there is no split that has a ratio close enough to the fixed one (\eg because the author worked only in a specific part of the project), we remove the author from the dataset.


We created two datasets by applying the described algorithm to the method creations by the developers in two aforementioned groups. The file tree in the IntelliJ IDEA project has a depth of 12. Since the increase in similarity value from $d-1$ to $d$ affects only files with depth $d$ and higher, we vary $d$ only in the range $[1;9]$, which includes over 95\% of the files.

Because of the restrictions on the ratio of training samples, for some authors we could not find a suitable separation at every level. For this reason, we filter out samples from these authors. The filtering left us with 26 authors out of 50. Finally, we obtained a dataset consisting of about 348,000 samples by 26 authors split at 9 different levels. At all the levels we have the same set of authors and code snippets, with the only difference being the split between training and evaluation part of the dataset.


\subsubsection{\textbf{Dataset with separation in time}} \label{sec:dataset-time-separation}
These datasets are designed to investigate whether developers' coding practices change over time. The high-level idea is as follows: we pick a set of method creations from a project (IntelliJ IDEA in our case), sort them by time, split into ten folds, then train a model on one fold and evaluate it on the others.

More specifically, we gather all events of method creation generated by the developers in groups with different amounts of samples per developer. Then, we sort all the methods written by each author by creation time and divide them into ten buckets of equal size. To preserve the same distribution of the authors across the buckets, we do the division independently for each programmer. In the end, most of the buckets for each developer contain methods written over 4 to 12 months. We also evaluated an alternative approach of splitting all the methods simultaneously, but this approach resulted in adding too much noise to the data (in fact, some programmers had joined the project later and the model trained on earlier folds did not have any information about them).

The resulting datasets consist of 400,000 and 200,000 events of method creation split into ten equal folds. The events are sorted in time and the difference between the indices of their folds can be used as the temporal distance between them. The distribution of the authors and the number of samples is uniform for every fold.

\subsection{Benefits of the data collection technique}
The proposed data collection technique enables gathering datasets that have several major benefits over existing datasets and capturing some important effects that are specific to real-world data:
\begin{itemize}
	\item \textit{Smaller gap between work context of code written by different authors.} While different developers still tend to work in different parts of the codebase, naming conventions, internal utility libraries, and the general domain are the same for everyone, since all code originates from the same codebase.
	\item \textit{Large number of samples available per author.} Existing datasets mostly work with up to several hundred code fragments per author. In the IntelliJ IDEA dataset collected with our technique, 21 developers created more than ten thousand methods each. The ability to collect multiple contributions for a single author makes the resulting data suitable for studying more fine-grained aspects of authorship attribution, such as the effect of the changes in coding practices over time on attribution accuracy.
	\item \textit{Broader domain of application.} Since our data collection technique allows one to collect data from any Git project, it is possible to investigate cross-project or cross-domain authorship attribution. 
\end{itemize}
 
\section{Evaluation on Collected Datasets}\label{sec:evaluation-idea}

We evaluate both our models (described in \Cref{sec:models}) on the datasets that we created from IntelliJ IDEA source code using the technique described in \Cref{sec:data}. Also, we adapt the model by Caliskan \emph{et al.}~\cite{Caliskan2015} to work with Java code and refer to it as \emph{JCaliskan}. We compare \emph{JCaliskan} to our models on the new datasets.

We also tried to reimplement the model by Yang \emph{et al.}~\cite{Yang2017}, since the work lacks an open-source implementation. However, we failed to reproduce the results mentioned in the paper~\cite{MLinProgrammingAuthorshipDetection}. In our case, using particle swarm optimization to train the model led to no improvements and strong overfitting. Since features used in their paper resemble the work by Caliskan \emph{et al.}~\cite{Caliskan2015}, we used JCaliskan for comparison.

\subsection{Separated work contexts}\label{sec:contextsplit}

First, we work with separated work contexts (\Cref{sec:dataset-context-separation}). These datasets contain nine different training/evaluation splits of the same large pool of method creation events, labeled with the method's author. Each split is parameterized with a depth value, which indicates the maximum possible depth of the lowest common ancestor of file changes in the training and evaluation sets.

\begin{figure}[htp]
\centering
    \subfloat[44 authors, 2,000 to 10,000 samples per author]{
       \includegraphics[width=\linewidth]{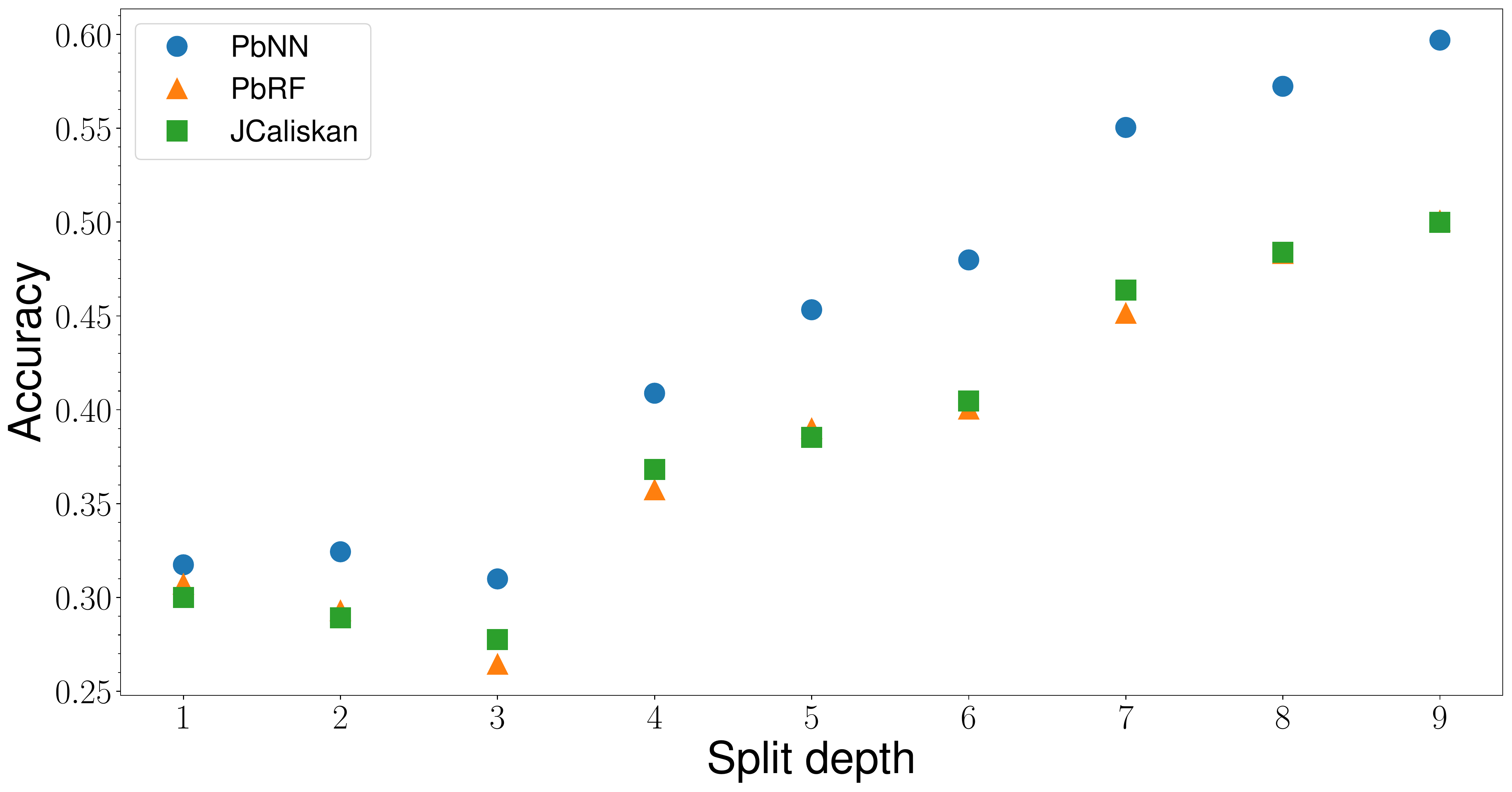}
    }
    
    \subfloat[21 authors, at least 10,000 samples per author]{
       \includegraphics[width=\linewidth]{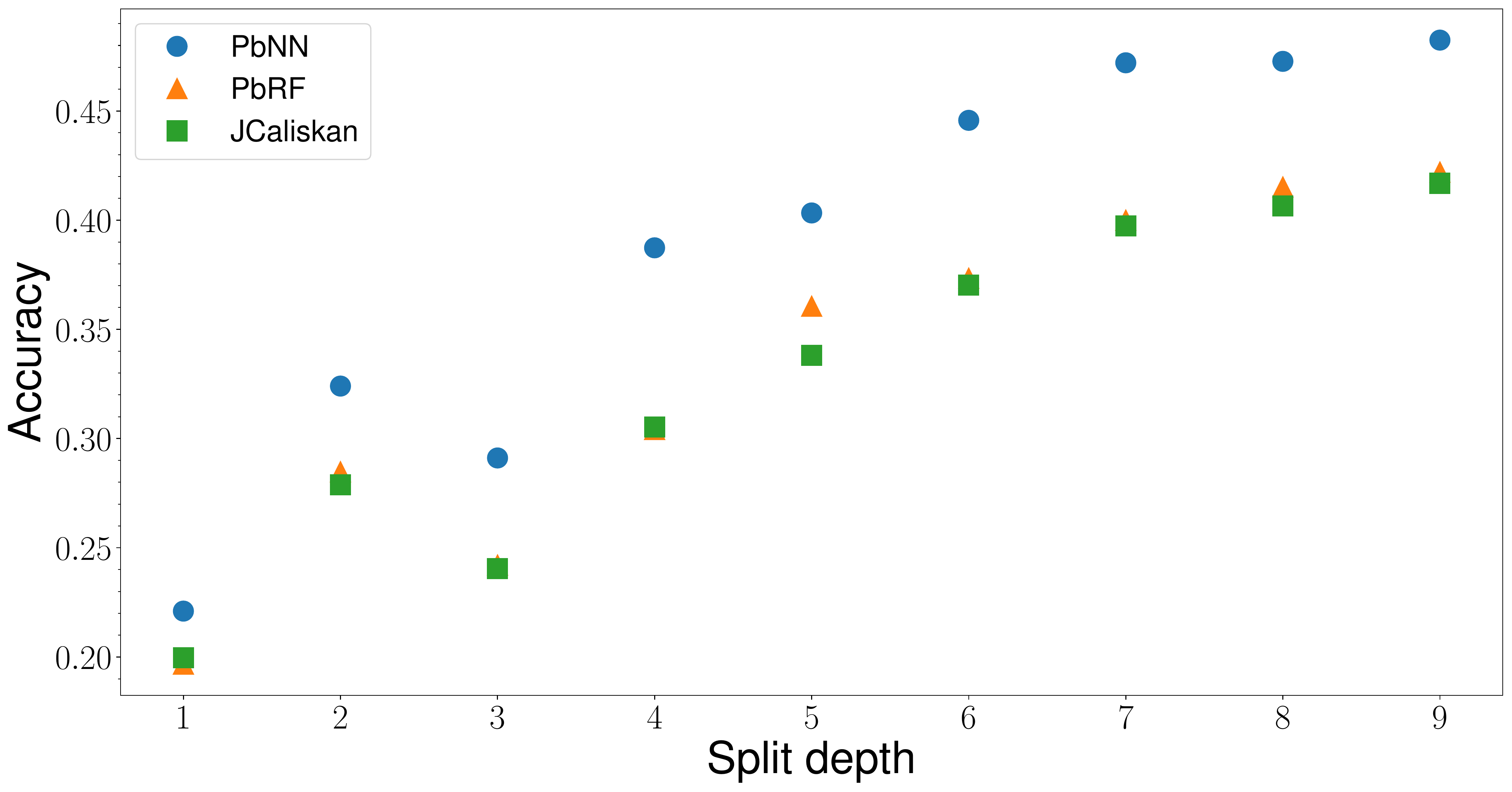}
    }
    \centering
    \caption{Models' accuracy on the datasets with separation of work context.}
    \label{fig:contextsplit}
\end{figure}

If the model is sensitive to the influence of work context, it should perform better for higher split depths, because with the growth of the split depth, training and testing sets become closer from the point of work context.
\Cref{fig:contextsplit} shows the dependency of the accuracy values of all three models on the split depth. Since the number of available samples per author is high and sufficient for proper training, the neural network model (PbNN) outperforms both random forest models (PbRF and JCaliskan) for each depth split and for both developer groups.

\Cref{fig:contextsplit} shows that the accuracy values increase with the depth, and at small depths the accuracy is much lower than in previous experiments (see \Cref{tab:results-comparison}). We tried to eliminate possible reasons for that, except for the difference in work context: the experiments were held on the same data points, the sizes of the training sets vary by less than 3\%, and we train models until convergence.

Thus, we conclude that \emph{work context strongly affects the accuracy of authorship attribution.}

\subsection{Time-separated dataset}\label{sec:timesplit}

To see if developers' coding practices change in time, which might affect the accuracy of authorship attribution models, we evaluate our models on two collected datasets with folds separated in time (\Cref{sec:dataset-time-separation}). The datasets contain samples of method creation from developers who did between 2,000 and 10,000 method creations and developers who did at least 10,000 method creations in the IntelliJ IDEA project. For each of the developers, the data has been divided by time into ten folds of equal size. This way we preserved the distribution of authors across folds, eliminating all differences between folds, except for the time when they were written.

We train a separate model on data from each fold except for the last. Then the models are tested on code fragments from subsequent folds. Thus, for ten folds we get nine trained models and 45 fold predictions. We expect a lower accuracy for more distant folds, if the developers' practices change in time.

\begin{figure}[htp]
\centering
    \subfloat[PbNN's accuracy]{
       \includegraphics[width=\linewidth]{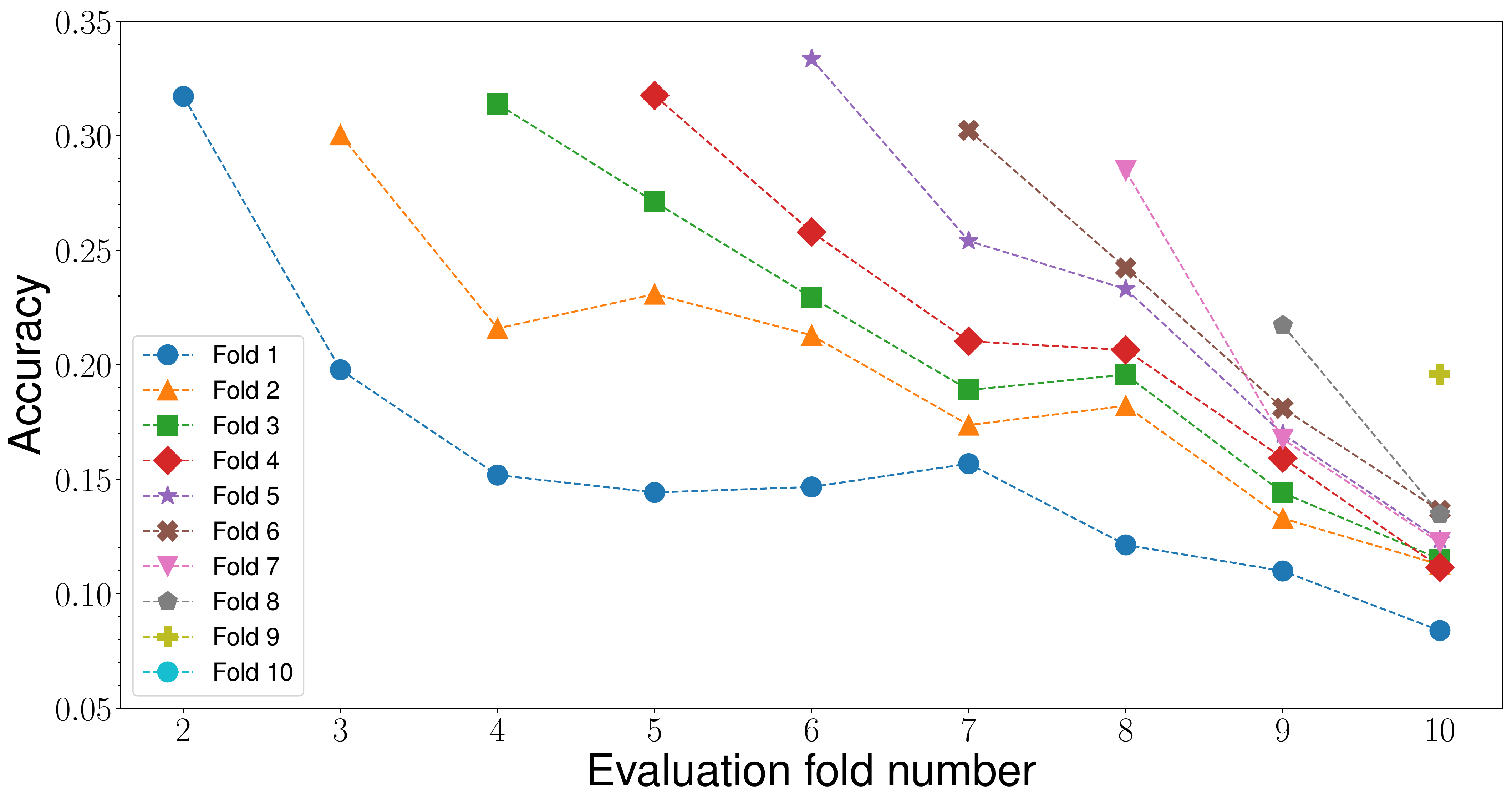}
    }
    
    \subfloat[PbRF's accuracy]{
       \includegraphics[width=\linewidth]{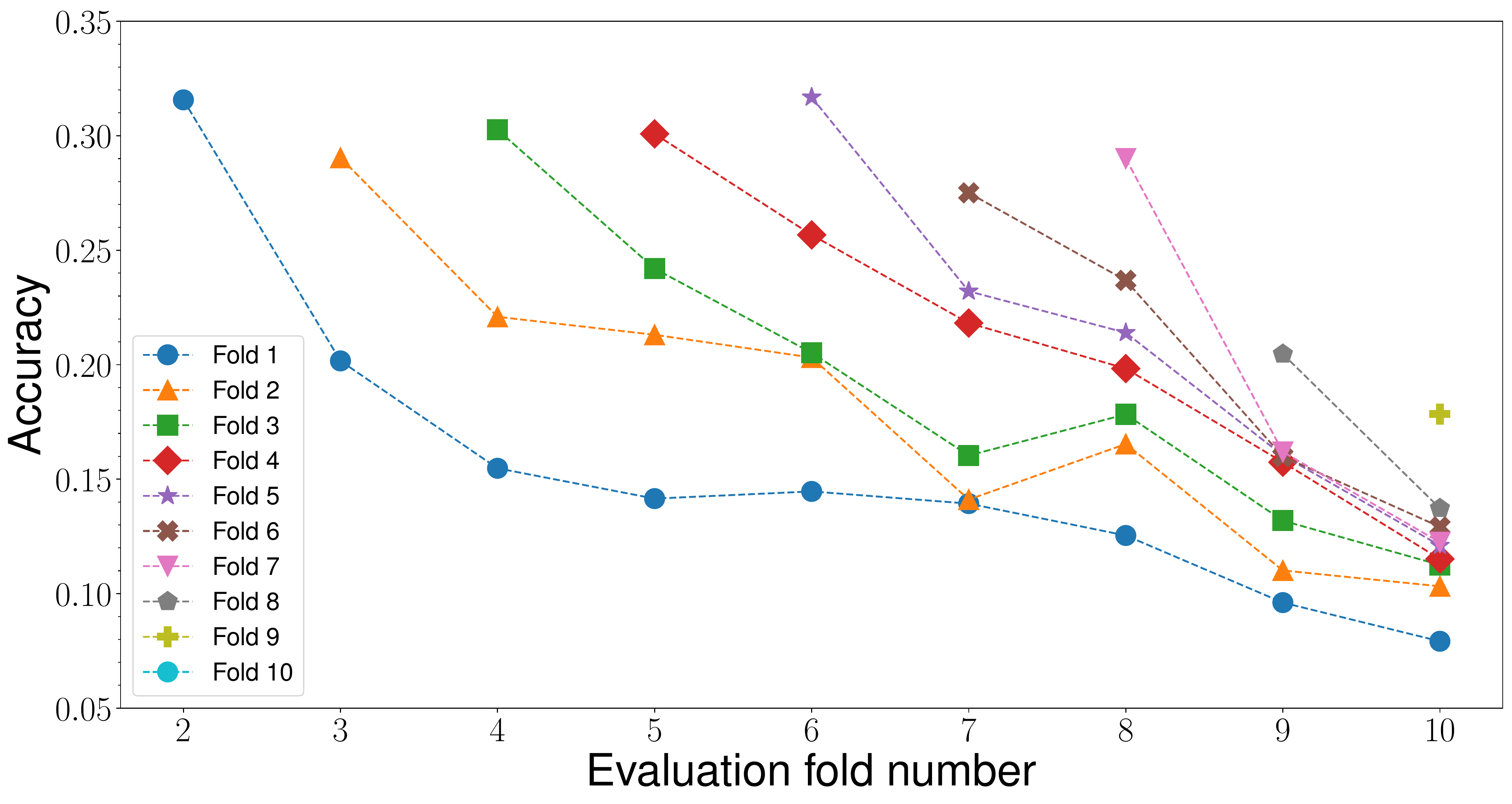}
    }
    
    \subfloat[JCaliskan's accuracy]{
       \includegraphics[width=\linewidth]{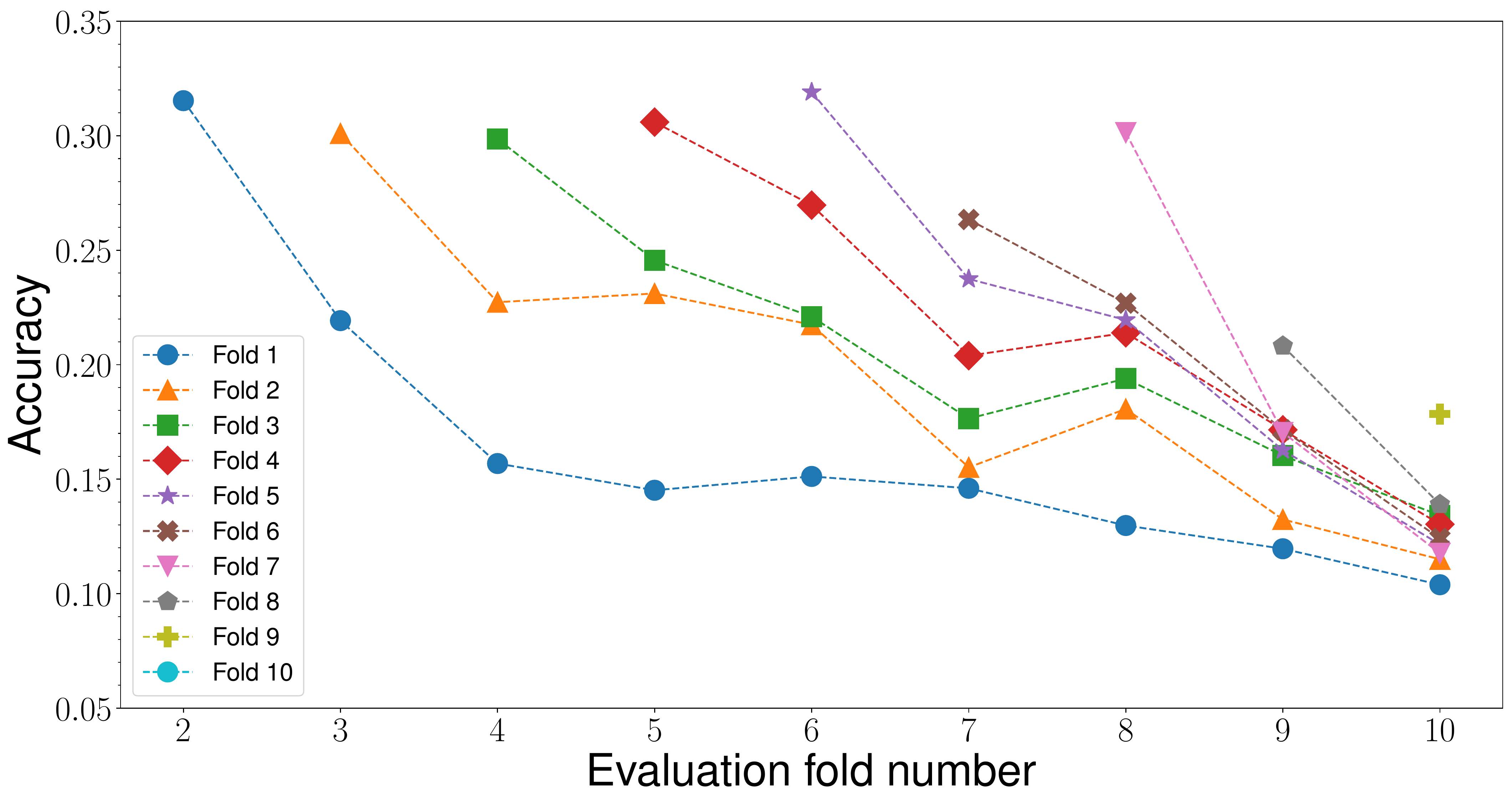}
    }
    \centering
    \caption{Models' accuracy on the dataset with separation in time (21 developers, at least 10,000 samples each). Lines are drawn for better readability and do not denote linear approximation.}
    \label{fig:timesplit}
\end{figure}

\begin{figure}[htp]
\centering
    \subfloat[PbNN's accuracy]{
       \includegraphics[width=\linewidth]{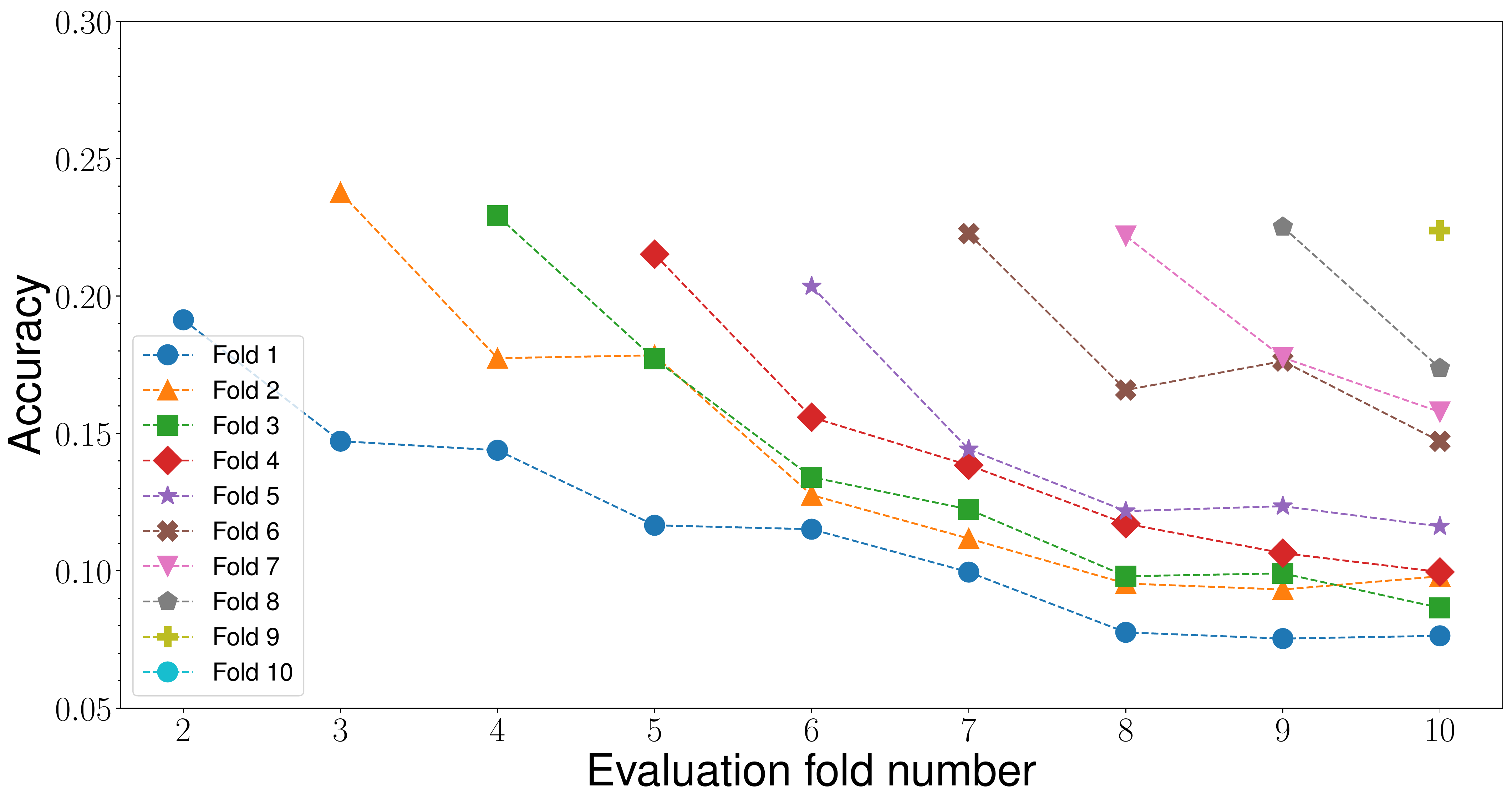}
    }
    
    \subfloat[PbRF's accuracy]{
       \includegraphics[width=\linewidth]{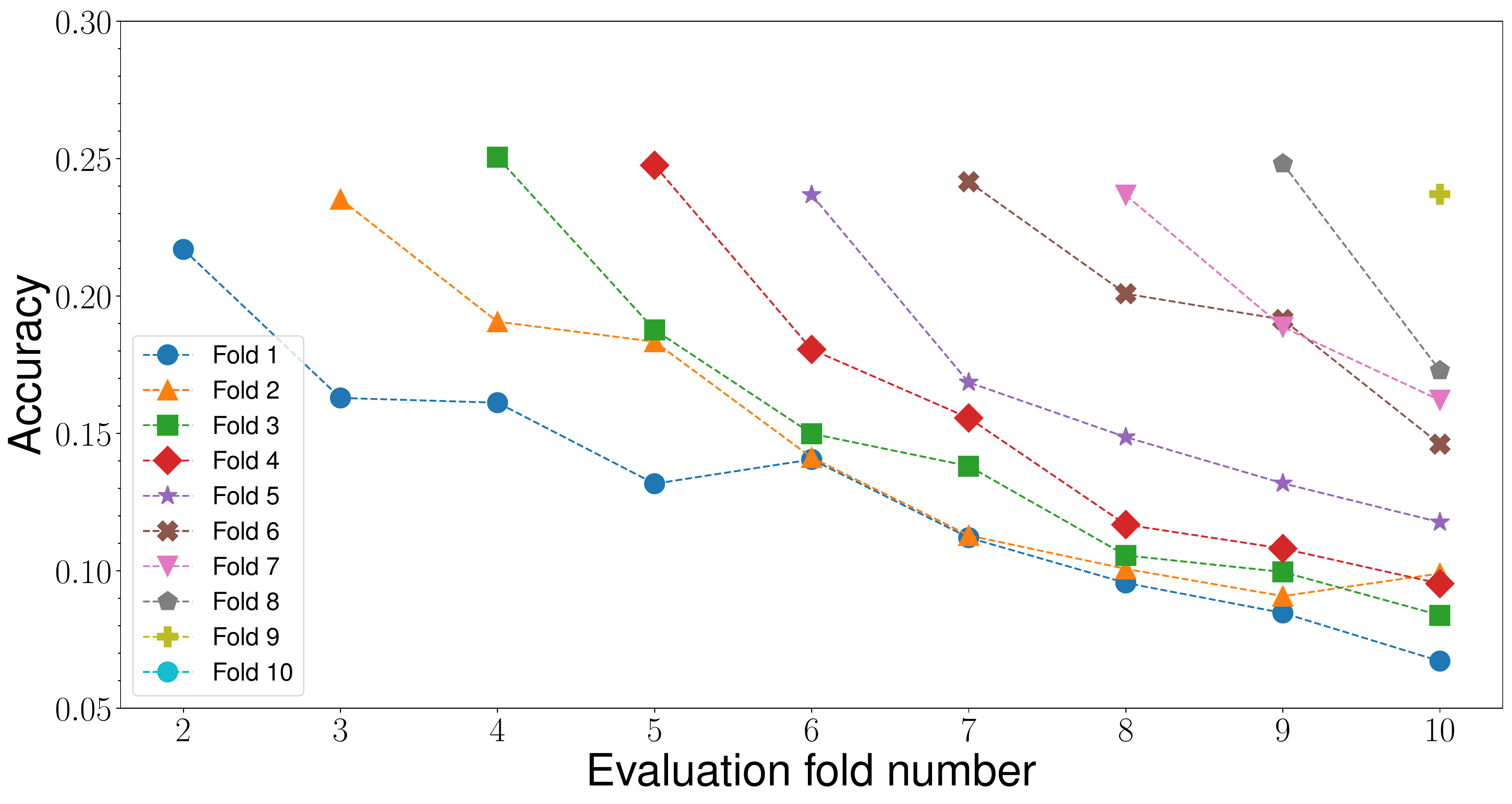}
    }
    
    \subfloat[JCaliskan's accuracy]{
       \includegraphics[width=\linewidth]{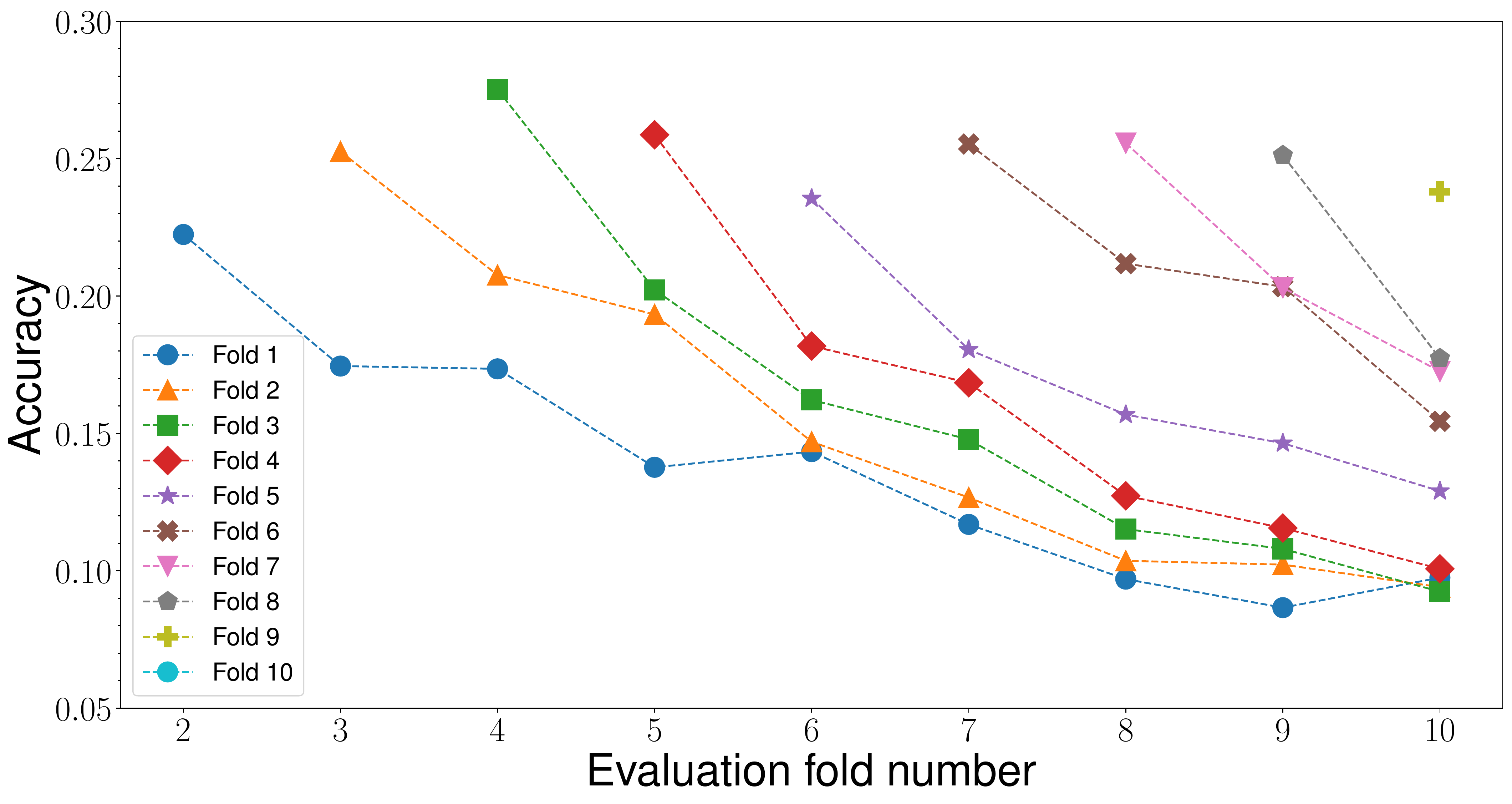}
    }
    \centering
    \caption{Models' accuracy on the dataset with separation in time (44 developers, 2,000 to 10,000 samples each). Lines are drawn for better readability and do not denote linear approximation.}
    \label{fig:timesplit2}
\end{figure}

The results for the models are presented in \Cref{fig:timesplit} and  \Cref{fig:timesplit2}. The neural network (PbNN) performs on par with random forest models (PbRF and JCaliskan) for the developers with at least 10,000 samples, but falls behind when the number of samples per author decreases. The graphs show that the accuracy of all models drops as distance in time grows, which confirms our hypothesis: \emph{evolution of coding practices affects the accuracy of authorship attribution.}

\subsection{Evaluation on other projects}

We also held a preliminary evaluation on two other large Java projects: Gradle~\cite{Gradle} and mule~\cite{Mule}. We applied the proposed data collection technique and created context-separated and time-separated datasets from each project. The results are consistent with the evaluation on IntelliJ IDEA:
\begin{itemize}
    \item Models' accuracy rapidly drops as we gradually separate contexts.
    \item Models' accuracy strongly depends on the distance in time between training and evaluation folds.
    \item PbNN outperforms both PbRF and JCaliskan when the number of available samples per author is high.
\end{itemize}

The detailed results and graphs are available in supplementary materials.


    

\section{Discussion}\label{sec:discussion}

Based on the evaluation results on the collected datasets, we conclude that both the difference in work contexts between training and testing data and the evolution of developers' coding practices have strong influence on the accuracy of authorship attribution models. However, our results are limited to Java language and several large open-source projects and need further research.

\subsection{Influence of The Work Context}

In \Cref{sec:evaluation-idea}, we described an experiment with separation of the work context between training and evaluation sets. We demonstrated that the model's accuracy decreases as we train and evaluate it on more distant (in terms of the codebase's file tree) pieces of code for each author. Specifically, the accuracy might vary by almost two times as we divide the same samples differently (\Cref{fig:contextsplit}).

We conclude that the gap between datasets used in previous work and the data observed in practical tasks is not negligible. Specifically, a model's accuracy can drop from 98\% in one setting (\Cref{tab:results-comparison}) to 22\% in another (\Cref{fig:contextsplit}). This suggests that, for evaluation to provide realistic information about the potential accuracy of a solution on data from conventional software engineering projects, researchers should use datasets where training and testing parts for each author belong to different environments.

The proposed dataset with gradual separation of training and evaluation sets can be used by researchers in future to measure their models' tendency to rely on context-related features rather than individual developers' properties.
The models we evaluated turned out to have a strong dependency on work context as well, with the accuracy dropping from 48\% to 22\% (PbNN) and from 42.2\% to 19.6\% (PbRF) for splits at depth 9 and 1. To lower the influence of work context on the model's accuracy, researchers could design context-independent features or add regularization terms.

\subsection{Evolution of Developers' Coding Practices}

Evaluation on the dataset with samples of code from each developer split in time showed that, as programmers' coding practices evolve over time, learning on older contributions to attribute authorship of the new code leads to a lower accuracy of attribution on a span of several months to years. To maximize the accuracy in potential real-world scenarios, we should use training data that is as relevant as possible, and re-train or fine-tune the models as we gather new data samples.

\subsection{Threats to Validity}

The proposed technique of data collection does not take into account automatically generated code, boilerplate code, or code embedded by an IDE. Its presence may bias the generated dataset. In order to mitigate this issue, the future work might employ techniques that detect such code and remove it from the dataset.

The collected dataset might contain code copied from third-party libraries, which was not completely written by the developer stated as the author in the VCS. Recent studies on open-source Java projects show that up to 20\% of methods can be copied from other sources~\cite{Golubev2020}, 6\% of Java projects are 100\% clones of other projects, and 14\% contain more than 50\% of cloned code~\cite{Lopes2017}. It means that for a repository that is not a clone, it is reasonable to expect that significantly fewer than 20\% of methods are copied from other places. The rest of the methods should be enough for the models to infer information about the authors. 

The experiment with gradual separation of work context relies on the proposed method to measure work context similarity as the depth of the lowest common ancestor in a file tree. Despite the provided rationale on why it is reasonable, for some projects the similarity of code might be weakly related to the location in the codebase due to specific codebase organization practices.

We found that the evolution of developers' coding practices strongly affects the accuracy of authorship attribution models. However, the observed drop in models' accuracy can be caused in part by the evolution of the whole project instead of the individual programmers. Also, the observed results are limited to the IntelliJ IDEA data.
To extend them to a general case, one needs to collect a dataset that comprises several projects with overlapping sets of developers, with projects divided between training and testing sets.




\section{Conclusion}\label{sec:conclusion}

Authorship attribution of source code has applications in software engineering in tasks related to software maintenance, software quality analysis, and plagiarism detection. While recent studies of authorship attribution report high accuracy values, they use language-dependent models and do not assess whether their datasets resemble data from real-world software projects.

We propose two models for authorship attribution of source code: PbNN (a neural network) and PbRF (a random forest model). Both models are language-agnostic and work with path-based representations of code which can be built for any syntactically correct code fragment. Our evaluation on datasets for C++, Python, and Java used in recent work shows that the suggested models are competitive with state-of-the-art solutions in terms of accuracy. In particular, they improve attribution accuracy on the Java dataset from 91.1\% to 97.9\% (PbNN) and 98.5\% (PbRF).

While demonstrating high accuracy, existing work in authorship attribution is evaluated on datasets that might be inaccurate in modeling real-world conditions. This may hinder their adoption in software engineering methods and tools.
We discuss a concept of \textit{work context}---the environment that influences the process of writing code---such as surrounding files, broader codebase, or team conventions. Taking work context into account, there is significant dissimilarity with previous results.
Another concern investigated in this work is the evolution of developers' coding practices and its potential impact on accuracy of authorship attribution. 

We suggest a novel approach to creation of authorship attribution datasets. In contrast to prior studies that are limited to projects with a single author, our approach can work with any Git project. We use our approach to process history of changes in a large Java project (IntelliJ IDEA Community repository on GitHub) and create several datasets to study the influence of work context and coding practices evolution on the accuracy of authorship attribution.

Evaluation of three models on the dataset with separation of work context shows that the accuracy goes down as similarity values decrease. As we gradually change the similarity level from maximal to minimal, the accuracy drops to the low of 22\%, which is much lower than 98\% achieved on the existing dataset of 40 single-authored projects.
For the experiment with folds divided in time, the accuracy drops as the time difference between training and testing folds increases, and the drop might also be significant: over 3 times for the most distant folds. We conclude that programmers' coding practices evolve over time, at least in large projects, and their evolution negatively affects quality of authorship attribution.

Our study demonstrates that existing solutions to authorship attribution can perform very differently when existing datasets are put into conditions close to real-world. This should be taken into account when evaluating authorship attribution approaches, especially during stages of data collection and training/testing division.

All the artifacts related to this work are publicly available on GitHub~\cite{JbrAuthorshipDetection} under MIT License.


\begin{acks}
Alberto Bacchelli gratefully acknowledges the support of the Swiss National Science Foundation through the SNSF Project \\
200021\_197227.
\end{acks}

\bibliographystyle{ACM-Reference-Format}
\bibliography{authorship-fse}

\end{document}